\documentclass[a4paper,11pt]{article}

\usepackage{graphicx}
\usepackage{algorithm}
\usepackage{algorithmicx}
\usepackage{algpseudocode}
\usepackage{amsmath}
\usepackage{epstopdf}
\usepackage{times}
\graphicspath{ {./}{figures/} }

\title{Count-Min Tree Sketch: Approximate counting for NLP}

\author{
	Guillaume Pitel, eXenSa\\
	\texttt{guillaume.pitel@exensa.com}
	\and
	Geoffroy Fouquier, eXenSa\\
	\texttt{geoffroy.fouquier@exensa.com}	
	\and
	Emmanuel Marchand, eXenSa\\
	\texttt{emmanuel.marchand@exensa.com}
	\and
	Abdul Mouhamadsultane, eXenSa\\
	\texttt{abdul.mouhamadsultane@exensa.com}
}

\begin{document}

\maketitle

\begin{abstract}
The Count-Min Sketch \cite{cormode2005improved} is a widely adopted structure for approximate event counting in large scale processing. In previous works \cite{pitel2015count,talbot2009succinct}, the original version of the Count-Min-Sketch (CMS) with conservative update has been improved using approximate counters \cite{morris1978counting,flajolet1985approximate} instead of linear counters. These structures are computationaly efficient and improve the average relative error (ARE) of a CMS at constant memory footprint. These improvements are well suited for NLP tasks, in which one is interested by the low-frequency items. However, if Log counters allow to improve ARE, they produce a residual error due to the approximation. In this paper, we propose the Count-Min Tree Sketch {\let\thefootnote\relax\footnote{Copyright \copyright 2016 eXenSa. All rights reserved.}} variant with pyramidal counters, which are focused toward taking advantage of the Zipfian distribution of text data.
\end{abstract}

\section{Approximate counting and NLP tasks}

The Count-Min Sketch \cite{cormode2005improved} (CMS) is a widely adopted structure for approximate event counting in large scale processing. With proved bounds in terms of mean absolute error and confidence, one can easily design a constant size sketch as an alternative to expensive exact counting for a setting where the total number of event types is approximately known.

The CMS is used in many applications, often with a focus on high frequency events \cite{cormode2005s}. However, in the domain of text-mining, highest frequency events are often of low interest: frequent words are often grammatical, highly polysemous or without any interesting semantics, while low-frequency words are more relevant. As a matter of fact, a common regularizations in text-mining consist in computing the Pointwise Mutual Information \cite{xu2007study} (equation \ref{eqn:PMI}) or Log-likelihood Ratio \cite{dunning1993accurate} between two words in order to estimate the importance of their cooccurrence. 
 
 \begin{subequations}
 	\label{eqn:PMI}
 	\begin{align}
 	pmi_i,j = \log \frac{p(i,j)}{p(i)p(j)}
 	\end{align}
 \end{subequations}

where $p(i)$ the probability to find word $i$ in the corpus and $p(i,j)$ is the probability that both words $i$ and $j$ appear in the same cooccurrence window.

This formula shows that higher frequency words will induce a relatively lower value at the end. Moreover, they all use a logarithm, hinting that only the order of magnitude is important. Another aspect of text-mining tasks, is that the distribution of word occurrence counts follows the Zipf law, which is highly skewed: most of the words have very few occurrences.

\section{Background on approximate counting}

The original version of the Count-Min-Sketch (CMS) is focused toward minimizing the absolute error and for uniform value distributions. A variant of CMS \cite{goyal2012sketch} has been proposed to compensate for the high relative error for low-frequency events, but the solutions explored tend to correct the errors instead of preventing them.

The Count-Min Log proposed by \cite{pitel2015count} improves the quality of the approximation by focusing on the relative error, while being highly efficient implementation-wise, but it does not take advantage of the Zipfian distribution of counts in text-mining data. TOMB structures \cite{van2009probabilistic}, on the other hand, combines logarithmic counters and multi-level Spectral Bloom Filters, and thus takes advantage of the Zipfian distribution. The TOMB structures are, however, not very efficient from a computational perspective, and they use MAX values as guards to use an upper level of counters.

\section{Count-Min Tree Sketch}

\begin{figure} \centering
	\includegraphics[width=0.6\linewidth]{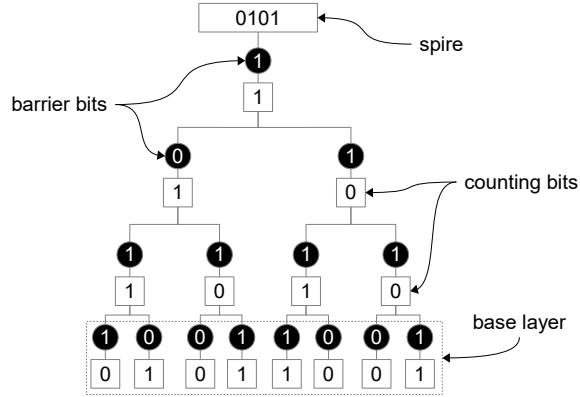}
	\caption[Count-Min Tree Sketches]{An example of CMTS with 4 layers and a spire.}
	\label{fig:hierarchie}
\end{figure}

We have designed and implemented the \textit{Count-Min Tree Sketch} (CMTS) \ref{fig:hierarchie}, a tree structure mixing \textbf{counting bits} and \textbf{barrier bits}. A counter starts at the bottom with a counting bit, if its first barrier bit equals one, then the next counting bit is also considered otherwise the counter just uses one counting bit. The counter stops at the bit right after the last contiguously activated barrier. A barrier bit, once set to $1$ can never be change anymore. At the very top, after the last barrier bit, a spire holds the remaining counting bits.

The general algorithms for updating and querying a Sketch is unchanged. However, given the complexity of the shared bits structure, the algorithms for getting the value, incrementing or merging the counters in case of distributed computing are much more complex than a Count-Min Sketch. Obtaining a competitive implementation has been a major endeavour.

\begin{figure} \centering
	\includegraphics[width=0.6\linewidth]{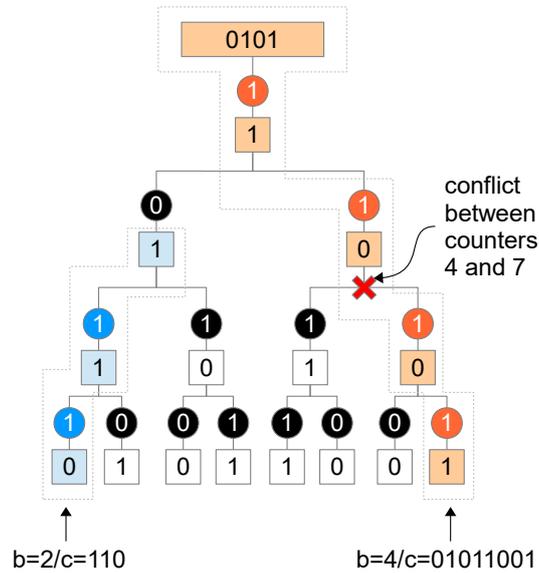}
	\caption[Count-Min Tree Sketches]{Getting the value of counters in a CMTS with 4 layers and a 4 bits spire. Real values of counters 0 and 7 are respectively 12 (6 from barrier, 6 from counting bits) and 119 (30 from barrier, 89 from counting bits). Counter 4 and 7 are in conflict and share values at the third level.}
	\label{fig:cmts_value}
\end{figure}

The principle of getting the value of a CMTS's counter is the following, and is illustrated figure \ref{fig:cmts_value}:
\begin{enumerate}
	\item The binary values of the barrier are gathered from the bottom cell of the counter, up to the first zero barrier. If there are 2 barrier bits contiguously set (like counter 0 in Figure \ref{fig:cmts_value}), then $b = 2$.
	\item ($b + 1$) value bits are gathered in counter $c$. If the bottom value bit is $0$ and the two others are $1$, then the counter is $c = 110_b = 6$
	\item The real value can finally be computed: $v = c + 2 \times \left ( 2^b-1 \right ) = 12$
\end{enumerate}

Incrementing can be implemented by getting the real value, incrementing it, then putting the result back into the CMTS, by setting the correct barrier and counting bits.

The principle of setting the value of a CMTS's counter is the following:
\begin{enumerate}
	\item The new barrier $nb$ is computed from the post increment value $nv$ using the following formula: $nb = min(nblayers, lsb(\frac{nv + 2}{4}))$, where $lsb(x)$ gives the bit length of $x$, i.e. the position of the first set bit plus one.\\ Within the previous example, the new value of counter 0 is $13$, $nblayer$ is $4$ and $lsb(\frac{13+2}{4}) = 2$. Hence, $nb = 2$
	\item The new counting bit is then computed by $nc = nv - 2 \times (2^{nb} - 1)$. Here $nc = 7, i.e., 111_b$
	\item Finally, the new barrier and the new value are set: $nb$ bits set to one for the barrier, and $nb + 1$ bits are set for the value using the new counting bit. \\In our example the number of barrier bits set to one remains the same ($2$), and $nb + 1 = 3$ bits are set for the value using the new counting bit $nc = 111_b$, thus the first three counters are set to $1$. In this case only the first one changes (from $0$ to $1$).
\end{enumerate}

The same basic get and set algorithms can be used for merging counters, even though many errors can be avoided by taking into account the possible overflows that can happen due to the conflicts in the tree structure.

\section{Empirical evaluation}

\subsection{Data}
We have verified this hypothesis empirically in the following setting: we count unigrams and bigrams of 140 million words of the English Wikipedia corpus. The small corpus analyzed contains 14.7 million distinct tokens. 

In the following, the ideal perfect count storage size corresponds, for a given number of elements, at the minimal amount of memory to store them perfectly, in an ideal setting. A high-pressure setting corresponds to a setting where the memory footprint is lower than the ideal perfect count storage size for the same number of elements. For our experiment, this size is 59 Megabytes (MiB).

All sketches use an implementation which is faster than a native dictionary structure (we use the native C++ STL implementation of an UnorderedMap). The memory footprint of this structure is approximately 815 MiB, i.e. 1380\% of the ideal perfect count storage size. 

\subsection{Variants}
We compare the estimates of four sketches: 
\begin{description}
\item[CMS-CU] is the classical linear Count-min Sketch with Conservative Update, 
\item[CMLS16-CU] is the Count-min-log Sketch with Conservative Update using a logarithmic base of 1.00025 and 16bits counters, and 
\item[CMLS8-CU] is the Count-min-log Sketch with Conservative Update using a logarithmic base of 1.08 and 8bits counters.
\item[CMTS-CU] is the Count-Min Tree Sketch with Conservative Update. Parameters are: 128 bits base, 32 bits spire.
\end{description}

\subsection{Error on counts}

\begin{figure*}
	\centering
	\includegraphics[width=0.9\linewidth]{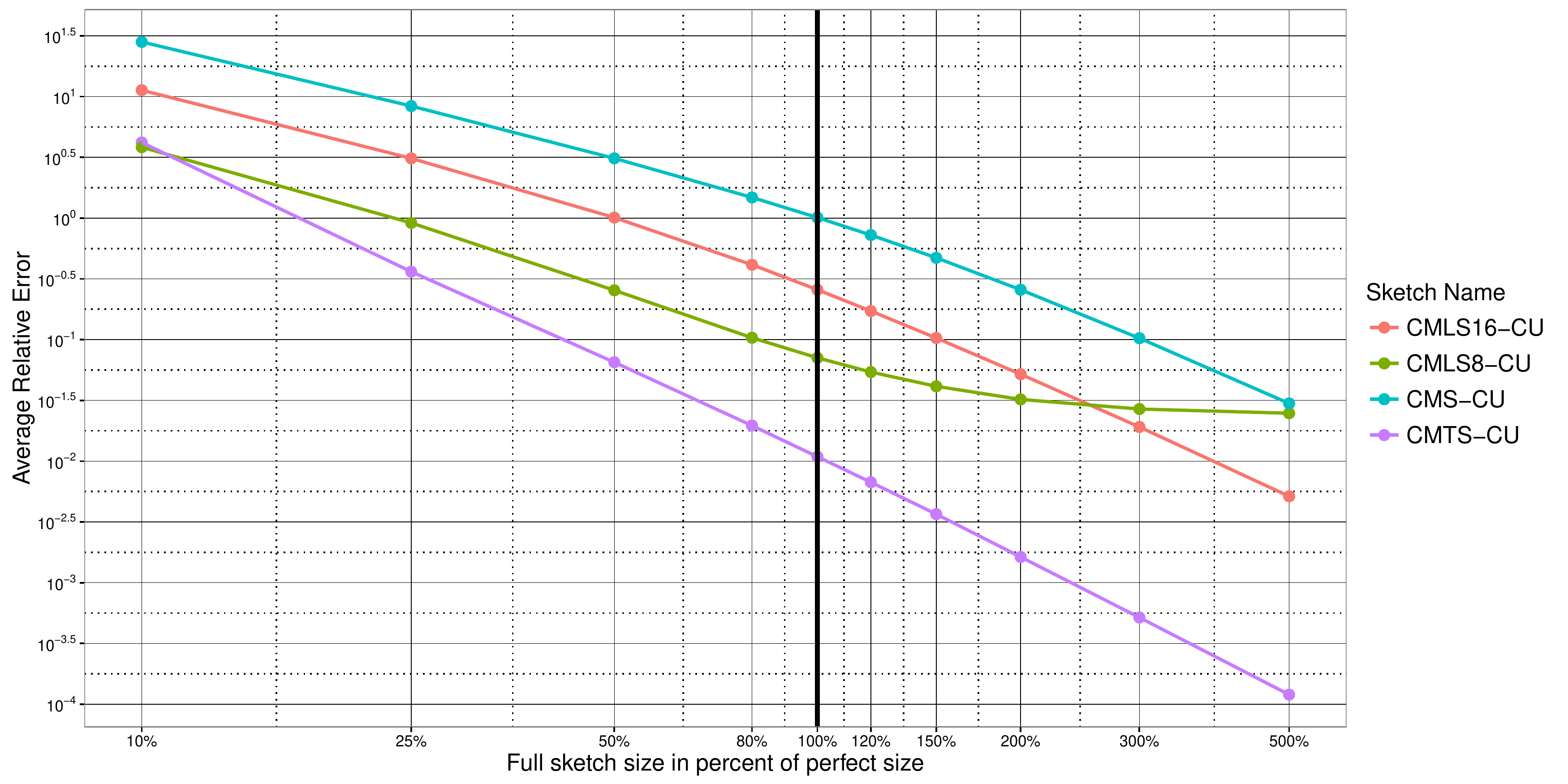}
	\caption[Average Relative Error of sketches counts]{Average Relative Error of estimated counts with Count-Min Sketch (CMS-CU, blue lines), Count-Min-Log 16bits (CMLS16-CU, red lines), Count-Min-Log 8bits (CMLS8-CU, green lines) and Count-Min Tree Sketch (CMTS-CU, purple). The bold vertical line corresponds to the ideal perfect counts storage size.}
	\label{fig:PYare}
\end{figure*}

\begin{figure*}
	\centering
	\includegraphics[width=0.9\linewidth]{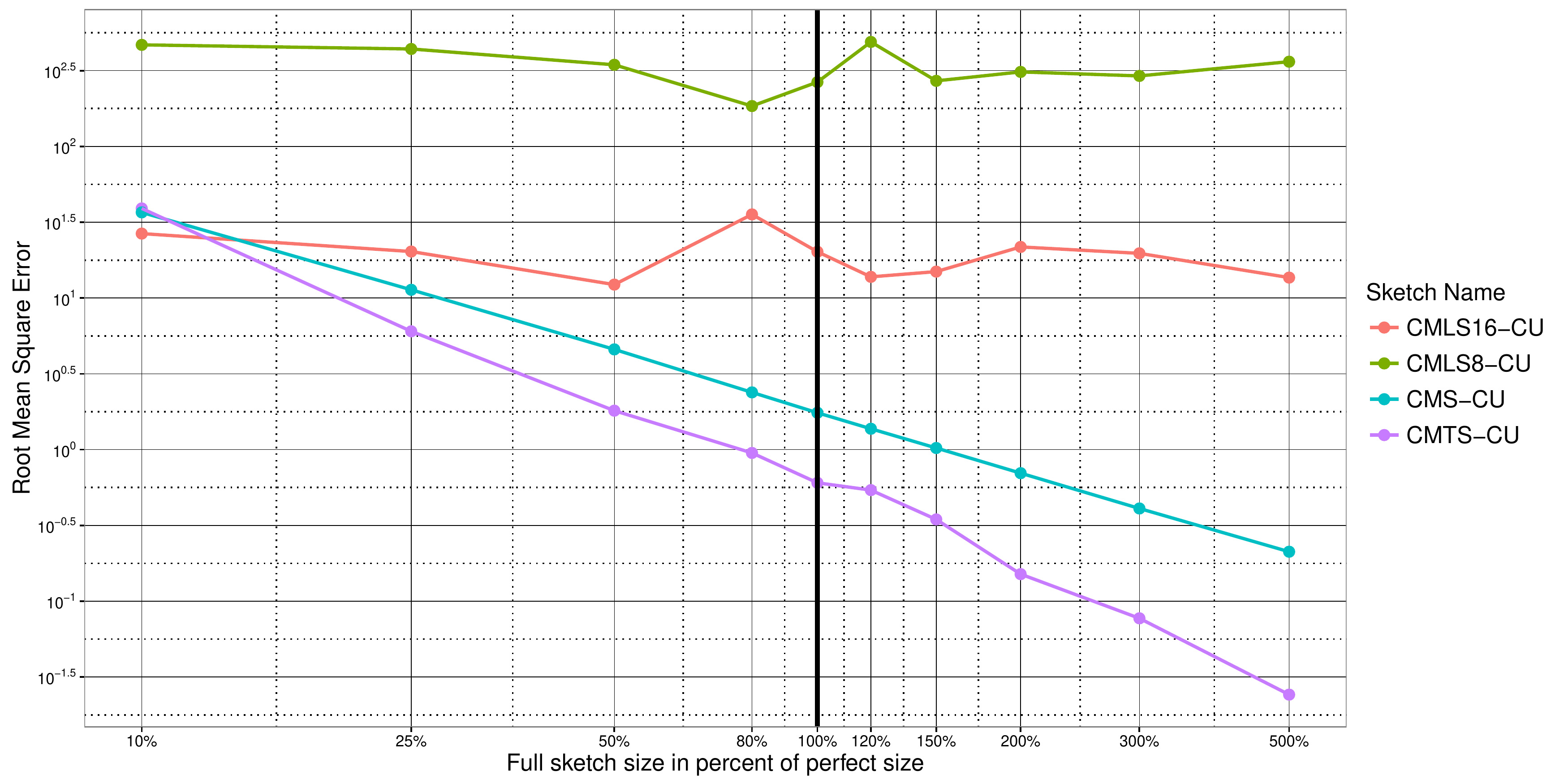}
	\caption[Average Relative Error of sketches counts]{Root Mean Square Error of estimated counts.}
	\label{fig:PYrmse}
\end{figure*}

\subsubsection{Relative Error}

The results for the Average Relative Error on simple counts of unigrams and bigrams are shown on figure \ref{fig:PYare}. The bold vertical line indicates the storage needed to memorize perfectly all the counts (the extra memory required for accessing the counters and storing the words is not taken into account).

This experiment shows that before the perfect storage mark, the estimation error of the CMLS16-CU is approximately 2 to 4 times lower than the error of the estimate of the CMS-CU. The CMLS8-CU error improvement over CMS-CU is in the range of 7 to 12 times. However, the CMLS8-CU reaches a minimal ARE of $10^{-1.5}$ at 200\% of the perfect size, and stops improving, due to the residual error caused by approximate counting. 

The CMTS produces an ARE of $10^{-2}$ at 100\% of the perfect size, and $10^{-3}$ at 300\% of the perfect size.

With respect to the standard CMS, CMTS improves the ARE by a factor of 100 at the perfect size. The size improvement for an equivalent error is approximately 800\%.

\subsubsection{Absolute Error}

The results for the Root Mean Square Error on simple counts of unigrams and bigrams are shown on figure \ref{fig:PYrmse}. Unlike logarithmic counters, which produce a high absolute error for high values, the CMTS-CU always performs better than the CMS-CU.

\subsection{Error on PMI}

\begin{figure*}
	\centering
	\includegraphics[width=0.9\linewidth]{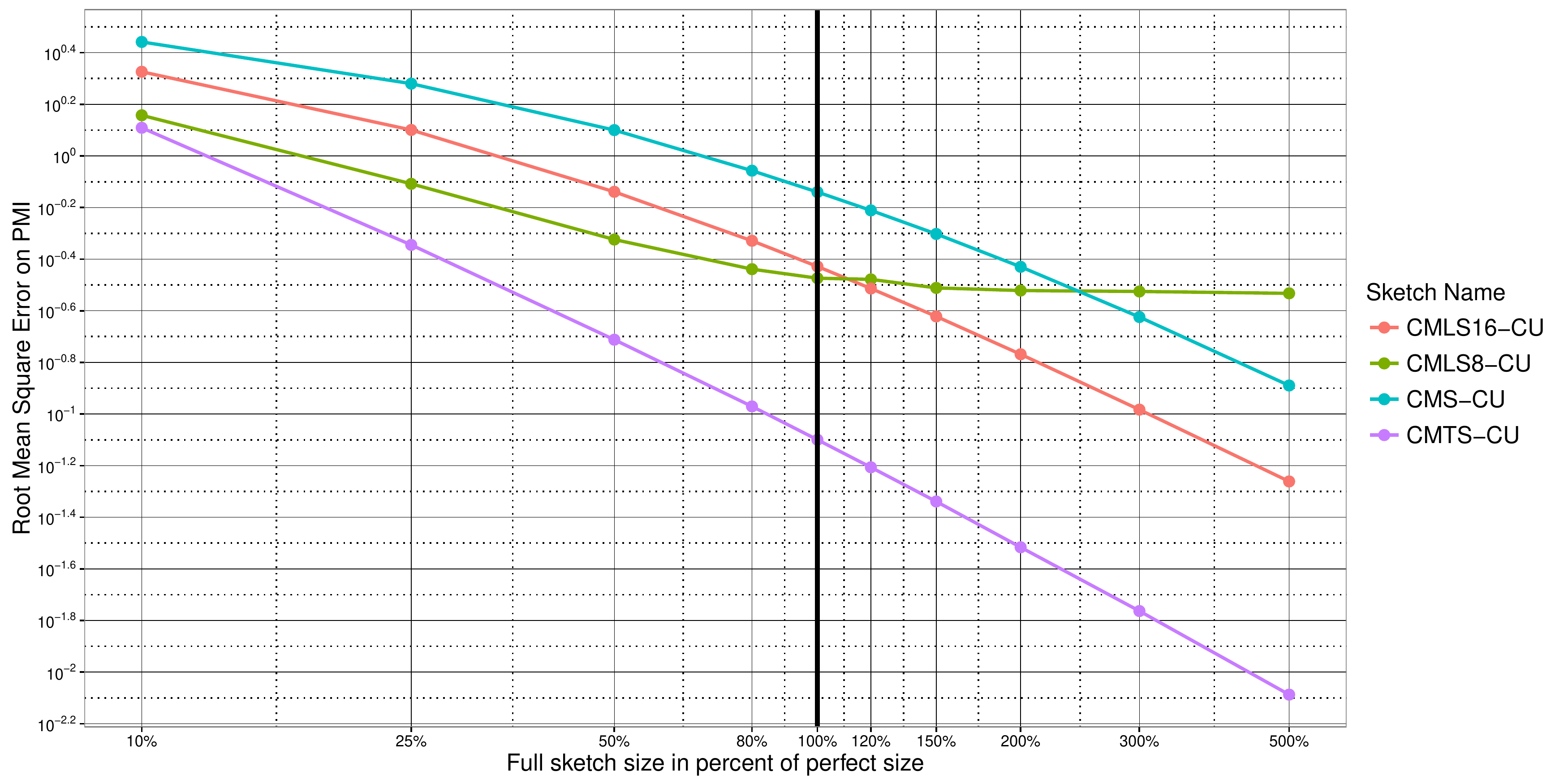}
	\caption{Root Mean Square Error of estimated PMI.}
	\label{fig:PYrmsepmi}
\end{figure*}

In a second step, we compute the Pointwise Mutual Information of the bigrams, and the error between the estimated PMI using counts from the sketch versus using the exact counts. The results for RMSE on estimated PMI are illustrated in figure \ref{fig:PYrmsepmi}.

These results show that, with sketches near the theoretical size of a perfect storage, both CMLS16-CU and CMLS8-CU outperforms CMS-CU by a factor of about 2 on the RMSE of the PMI. CMTS-CU outperforms CMS-CU by a factor of 10 for the RMSE at the perfect size, and the size improvement ratio is again approximately about 800\%.

\subsection{Behaviour under very high pressure}

When the memory footprint is less than 10\% of the ideal storage size, the CMTS performance degrades faster than all other variants. We consider this not to be a real problem for two reasons. First, at this pressure, the RMSE on the PMI is in the range $[1.5,2.5]$, and the ARE on counts is in the range $[4,31]$, which are too high for being practically useable.

\section{Conclusion and perspectives}

In this paper we show how approximate counting can take advantage of the Zipfian distribution of the textual data to improve precision to an unprecedented level: improvements over standard Count-Min Sketch range from 100 times smaller for the ARE, and more than 10 times smaller for the RMSE on the PMI.

The Count-Min Tree Sketch can be implemented in an efficient way. Although it's slightly less computationaly efficient than standard CMS, it remains competitive with native maps implementations.

Moreover, like for the Count-Min Sketch, an unsynchronized multithreaded implementation shows very little adversarial effect on the precision of most common sketch sizes: the effects on precision only appear at very low errors ($10^{-5}$).

\end{document}